\documentclass[aps,prb,reprint,groupedaddress]{revtex4-2}

\usepackage{amsmath}
\usepackage{amssymb}
\usepackage{hyperref}
\usepackage{epsfig}
\usepackage{graphicx}
\usepackage{subfigure}
\usepackage{latexsym}
\usepackage[dvipsnames]{xcolor}
\usepackage{fullpage}
\usepackage{epstopdf}
\usepackage{dcolumn}
\usepackage{bm}
\usepackage{ulem}
\usepackage{units}

\newcommand{\Nico}[1]{\textcolor{black}{#1}}
\newcommand{\new}[1]{\textcolor{black}{#1}}
\newcommand{\johann}[1]{\textcolor{black}{#1}}
\newcommand{\myparallel}{\mathbin{\!/\mkern-5mu/\!}}
 
\newcommand{\stkout}[1]{\ifmmode\text{\sout{\ensuremath{#1}}}\else\sout{#1}\fi}

\begin{document}

\title{\Nico{Ampère phase in frustrated magnets}}

\author{N. Rougemaille, J. Coraux, B. Canals} 
\affiliation{Université Grenoble Alpes, CNRS, Grenoble INP, Institut NEEL, 38000 Grenoble, France}

\date{\today}

\begin{abstract}
We report a new class of algebraic spin liquids, in which the macroscopically degenerate ground state manifold is not Coulombic, like in spin ices, but Ampère-like.
The local constraint characterizing an Ampère phase is not a Gauss law, but rather an Ampère law, i.e., a condition on the curl of the magnetization vector field and not on its divergence.
As a consequence,
the excitations evolving in such a manifold are not magnetically charged \textit{scalar} quasiparticles, the so-called magnetic monopoles in Coulomb phases, but instead \textit{vectorial} magnetic loops (or fictional current lines).
We demonstrate analytically that in a macroscopically degenerate manifold inheriting the properties of a cooperative paramagnet and subject to a local curl-free contraint, magnetic correlations decay in space with a power law whose exponent is the space dimension $d$: the Ampère phase is a $d$-algebraic spin liquid.
Using Monte Carlo simulations with appropriate cluster dynamics, we confirm this physics numerically in two- and three-dimensional examples, and illustrate how the Ampère phase compares to its Coulomb counterpart.
\end{abstract}

\maketitle

\section{Introduction}

Although elementary particles can be electrically charged, none of them has ever been found to carry a magnetic charge.
However, “the capacity of many-body systems to provide emergent mini-universes with vacua quite distinct from the one we inhabit" \cite{Moessner2016} allows the investigation of a wealth of phenomena in solid state physics, through the concept of \textit{quasiparticles}.
For example, certain highly frustrated magnets, such as spin ice compounds \cite{Harris1997, Gingras2011, Bramwell2013, Bramwell2020}, host excitations behaving as deconfined charged quasiparticles interacting via a Coulomb potential \cite{Ryzhkin2005, Castelnovo2008, Fennell2009, Jaubert2009, Jaubert2011, Castelnovo2012}.
Emerging in a purely spin system, these pseudo-charges are of magnetic origin, and can be seen as classical analogs of the elusive magnetic monopoles in high-energy physics.

The concept of magnetic monopoles is deeply rooted in the extensively degenerate ground state property of many highly frustrated magnets.
Indeed, describing such a disordered ground state manifold by a static coarse-grained vector field $\vec{\mathcal{F}}(\vec {r})$, and provided its statistics is similar to the one of a \textit{cooperative} paramagnet, Henley showed that the system free energy $F$ ressembles the energy density of a magnetic field \cite{Henley2010}:
\begin{equation}
\label{eqHenley}
\frac{F}{k_\mathrm{B} T} \propto \frac{K}{2} \int_{V_\mathrm{t}} \vec{\mathcal{F}}^2 (\vec{r})\; d^d \vec{r},
\end{equation}
where $K$ is a constant, $T$ the temperature, $k_\mathrm{B}$ the Boltzmann constant, $V_\mathrm{t}$ the total volume of the system, and $d$ the space dimension.
If now the physics of the frustrated magnet (or the considered model) is such that the vector field obeys a hard constraint, such that the sum of the fluxes at each lattice vertex is zero (divergence-free condition), like in spin ice compounds, then the correlations of $\vec{\mathcal{F}}$ have the functional form of a dipole-dipole interaction at large distances \cite{Henley2005}, and the analogy with electrostatics / magnetostatics is complete. 
Defects violating the local constraint then behave as effective charges, sources or sinks of the vector field $\vec{\mathcal{F}}$, and have been dubbed magnetic monopoles.
Such a disordered state of matter has been coined a Coulomb phase \cite{Henley2010}.

\section{Motivation}

Only systems or models having a divergence-based constraint have been considered so far \cite{Gingras2023}.
The fundamental question we address in this work is whether the analogy with electromagnetism and Maxwell equations can be further developed by considering the curl property of the vector field rather than its divergence. 
Generalizing Henley's approach developed for Coulomb phases, we show analytically that a curl-free constraint on the vector field $\vec{\mathcal{F}}(\vec {r})$ also gives rise to a $d$-algebraic spin liquid, which we coin an Ampère phase.
This analytical description is valid regardless of the system dimension \footnote{Of course, when the dimensionality of the curl is reduced, for instance for coplanar spins, the Coulomb and Ampère phases are locally equivalent. Nevertheless, we emphasize that this is related to the nature of the degree of freedom, and not to the space dimensionality. Furthermore, the properties of each phase may be quite different. In this work, we give 2D and 3D examples of Ampère phases that fragment. Their Coulomb counterparts, the 6-vertex model and the canonical pyrochlore spin ice, respectively, do not.}.
The properties of the Ampère phase is first illustrated considering the square lattice as a pedagogical case study, and we demonstrate the existence of a one-to-one correspondence between the Coulomb and Ampère phases on the two-dimensional (2D) square lattice.
This correspondance allows us to determine an Ising spin Hamiltonian characterized at low-energy by an Ampère spin liquid.
We then implement the curl-free contraint in a vertex model on the pyrochlore lattice to evidence an Ampère phase physics in three dimensions (3D).

\section{Algebraic magnetic correlations in the Coulomb and Ampère phases} 

\johann{In this section, we derive an analytical expression for the spin-spin correlations in the Coulomb and Ampère phases, and demonstrate they decay algebraically with the distance in both cases. 
This description follows the analysis provided by Henley in his seminal paper on Coulomb phases \cite{Henley2010}, but extends the concept to a local, curl-free contraint.
No assumption is made on the space dimensionality, so the result holds for 2D and 3D systems.
}

\johann{We consider a lattice dressed with a spin vector field associated to a cooperative paramagnet.
The manifold we are interested in is thus highly degenerate. 
}
\johann{We make the \textit{assumption} that the elementary bricks of the vector field, such as the vertices or plaquettes, eventually subjected to a local constraint, are independent and identically distributed random variables.
Without loss of generality, these variables distribute in a symmetric manner about a central value. 
Coarse-graining the spin vector field leads to $\vec{\mathcal{F}}(\vec {r})$, which extends over a certain volume $V$ within which it averages to zero. 
According to the central limit theorem, for large-enough cells, the statistics of the square of $\vec{\mathcal{F}}$ is well described by a centered Gaussian law,}

\begin{equation}
\label{eqN}
\mathcal{N}\left(\vec{\mathcal{F}}(\vec{r})\right)\propto \exp \left( -\frac{\vec{\mathcal{F}}  (\vec{r})^2}{2\sigma^2_{\vec{\mathcal{F}}}} \right).
\end{equation}

\johann{The systems of interest here are invariant by translation, meaning that fluctuations of the field do not depend on $\vec{r}$, and $\sigma^2_{\vec{\mathcal{F}}}=\sigma$ is a constant. 
The central limit theorem stipulates that the variance is proportional to the number of bricks, i.e., proportional to $V$, such that, assuming the coarse-graining is uniform, $\sigma^2=V/K$, with $K$ a coefficient that is specific to the considered system.}

\johann{In the considered cooperative paramagnet, the free energy is $F = E - TS$, where $E$ is the ground state energy, $T$ the temperature, and $S$ the entropy of the manifold.
Because $E$ is constant in the temperature range we consider, the energy plays no role, and the free energy is entropy-driven.
Defining an entropy function of the space coordinate, $\mathcal{S}_V(\vec{r})$ evaluated on a volume $V$ around $\vec{r}$, and introducing the entropy density $s(\vec{r})=\mathcal{S}_V(\vec{r})/V$, the free energy $F_V(\vec{r})$ at point $\vec{r}$ writes}

\begin{equation}
\label{eqFlocal}
\frac{F_V(\vec{r})}{T} = - \int_V s(\vec{r}) d^d \vec{r}.
\end{equation}

\noindent \johann{From Eq.~\ref{eqN}, it follows that}

\begin{equation}
\label{eqs}
s(\vec{r})=\frac{k_\mathrm{B}}{V} \log \left(\mathcal{N}\left(\vec{\mathcal{F}}(\vec{r})\right)\right) = s_0 - \frac{k_\mathrm{B} K \vec{\mathcal{F}}(\vec{r})^2}{2}.
\end{equation}

\noindent \johann{For the full system, as already noted in the introduction, the free energy resembles the energy density of a magnetic field}

\begin{equation}
\label{eqFtotal}
\frac{F}{k_\mathrm{B}T}\propto\frac{K}{2} \sum_{\vec{r}} \vec{\mathcal{F}}(\vec{r})^2,
\end{equation}

\noindent
\johann{where we have performed a discrete sum over the contributions of the cells centered at the $\vec{r}$ positions (rather than a continuous summation as in Eq.~\ref{eqHenley}). 
We may now expand the vector field on its $\vec{q},{\vec{q}\,}'$ Fourier components,}

\begin{equation}
\label{eqFtotalFourier}
\frac{F}{k_\mathrm{B}T} \propto \frac{K}{2N} \sum_{\vec{r}}\sum_{\vec{q}}\sum_{{\vec{q}\,}'} e^{i \vec{q}\cdot\vec{r}} e^{i {\vec{q}\,}'\cdot\vec{r}} \vec{\mathcal{F}}_{\vec{q}} \cdot \vec{\mathcal{F}}_{{\vec{q}\,}'}.
\end{equation}

\noindent \johann{Using $\sum_{\vec{r}} e^{i \vec{q}\cdot\vec{r}} e^{i {\vec{q}\,}'\cdot\vec{r}} = N \delta_{\vec{q}+{\vec{q}\,}'}$, and $\vec{\mathcal{F}}_{-\vec{q}} = \vec{\mathcal{F}}^*_{\vec{q}}$ (the vector field is real),}
\begin{equation}
\label{eqFtotalFourierFinal}
\frac{F}{k_\mathrm{B}T} \propto \frac{K}{2} \sum_{\vec{q}} \vec{\mathcal{F}}^2_{\vec{q}}.
\end{equation}

\noindent \johann{We hence deal, also in reciprocal space, with a set of independent Gaussian random variables.}

\johann{We stress that no hypothesis was necessary so far about $\vec{\mathcal{F}}(\vec{q})$. We only assumed that all configurations of the ground state manifold have the same energy, and that each configuration is well described by a sum of independent and identically distributed random variables. Within this framework, $\left\langle \mathcal{F}(\vec{q}) \cdot \mathcal{F}({\vec{q}\,}') \right\rangle = \delta_{\vec{q}+{\vec{q}\,}'}/K$ whatever the dimension of the system (no volume appears here due to integration in reciprocal space), which is the analog of the autocorrelation in real space.}

\johann{We now consider two particular directions, parallel ($\parallel$) and perpendicular ($\perp$) to the $\vec{q}$ vector. The vector field's corresponding components write $\vec{\mathcal{F}}_\perp(\vec{q})=\vec{\mathcal{F}}(\vec{q})-[\hat{q}\cdot \vec{\mathcal{F}}(\vec{q})]\hat{q}$ and $\vec{\mathcal{F}}_\parallel(\vec{q})=[\hat{q}\cdot\mathcal{F}(\vec{q})]\hat{q}$, with $\hat{q}=\vec{q}/|q|$. After some algebra:}

\begin{equation}
\label{eqCorrelFperppara}
\begin{cases}
\left\langle \vec{\mathcal{F}}^\mu_\perp (\vec{q}) \cdot \vec{\mathcal{F}}^\nu_\perp (\vec{q}) \right\rangle & = \frac{1}{K} \left( \delta_{\mu\nu} -\frac{q_\mu q_\nu}{q^2}\right) \\
\left\langle \vec{\mathcal{F}}^\mu_\parallel (\vec{q}) \cdot \vec{\mathcal{F}}^\nu_\parallel (\vec{q}) \right\rangle & = \frac{1}{K} \frac{q_\mu q_\nu}{q^2},
\end{cases}
\end{equation}

\noindent
\johann{with $\mu,\nu$ denoting space components. These expressions show that in the absence of specific constraint, the field-field correlations are \textit{not} algebraic, but constant. $\vec{\mathcal{F}}$ actually describes a true paramagnet (an uncorrelated manifold). However, the transverse and longitudinal components are, \textit{separately}, algebraic.
As we will see below, this result has important consequences, especially to understand the difference between a Coulomb phase and its Ampère counterpart.}

\johann{We now discuss the effect of a local constraint induced by the physics of the manifold. 
The constraint considered by Henley \cite{Henley2010} is that the spin variables satisfy a zero divergence condition on the lattice vertices. 
In reciprocal space, this translates as $\vec{q}\cdot\vec{\mathcal{F}}(\vec{q})=0$, so $\vec{\mathcal{F}}(\vec{q})=\vec{\mathcal{F}}_\perp(\vec{q})$. In this case, Eq.~\ref{eqCorrelFperppara} simplifies to}

\begin{equation}
\label{eqZeroDivCorrel}
\begin{split}
\left\langle \vec{\mathcal{F}}^\mu (\vec{q}) \cdot \vec{\mathcal{F}}^\nu (\vec{q}) \right\rangle & = \left\langle \vec{\mathcal{F}}^\mu_\perp (\vec{q}) \cdot \vec{\mathcal{F}}^\nu_\perp (\vec{q}) \right\rangle \\
& = \frac{1}{K} \left( \delta_{\mu\nu} -\frac{q_\mu q_\nu}{q^2}\right).
\end{split}
\end{equation}

\noindent This is the Coulomb phase introduced by Henley and the correlations of the corse-grained vector filed are algebraic. 
\johann{Key for this work, we now consider a curl-free constraint, corresponding to $\vec{q}\times\vec{\mathcal{F}}(\vec{q})=\vec{0}$ in reciprocal space. 
Under such a condition, Eq.~\ref{eqCorrelFperppara} simplifies to}

\begin{equation}
\label{eqZeroCurlCorrel}
\begin{split}
\left\langle \vec{\mathcal{F}}^\mu (\vec{q}) \cdot \vec{\mathcal{F}}^\nu (\vec{q}) \right\rangle & = \left\langle \vec{\mathcal{F}}^\mu_\parallel (\vec{q}) \cdot \vec{\mathcal{F}}^\nu_\parallel (\vec{q}) \right\rangle \\
& = \frac{1}{K} \frac{q_\mu q_\nu}{q^2}.
\end{split}
\end{equation}

To extend the parallel with Maxwell's equations, we call a phase obeying this constraint an Ampère phase, within which magnetic correlations are algebraic. 
Remarkably, the algebraic correlations of the Ampère phase are complementary to those of the Coulomb phase, such that $\left< \vec{\mathcal{F}}_{\perp}^\mu (\vec{q}) \; \vec{\mathcal{F}}_{\perp}^\nu (\vec{q}) \right> + \left< \vec{\mathcal{F}}_{\myparallel}^\mu (\vec{q}) \; \vec{\mathcal{F}}_{\myparallel}^\nu (\vec{q}) \right> = \delta_{\mu\nu}/K $.
In other words, a pure paramagnet can be viewed as a sum of two algebraic spin liquids, within which the associated coarse-grained vector field is constrained to be divergence-free (the Coulomb phase) and curl-free (the Ampère phase).

Topological defects in the Ampère phase are thus sinks and sources of magnetization curl.
They are not a \textit{scalar} quantity, like magnetic monopoles, but rather a \textit{vectorial} circulation.
The Gauss theorem used to describe monopoles in the Coulomb phase must be replaced by the Ampère theorem: to the magnetization circulation, one can associate a fictional current line perpendicular to the lattice plane.

\begin{figure}[ht!]
\includegraphics[width=7.6 cm]{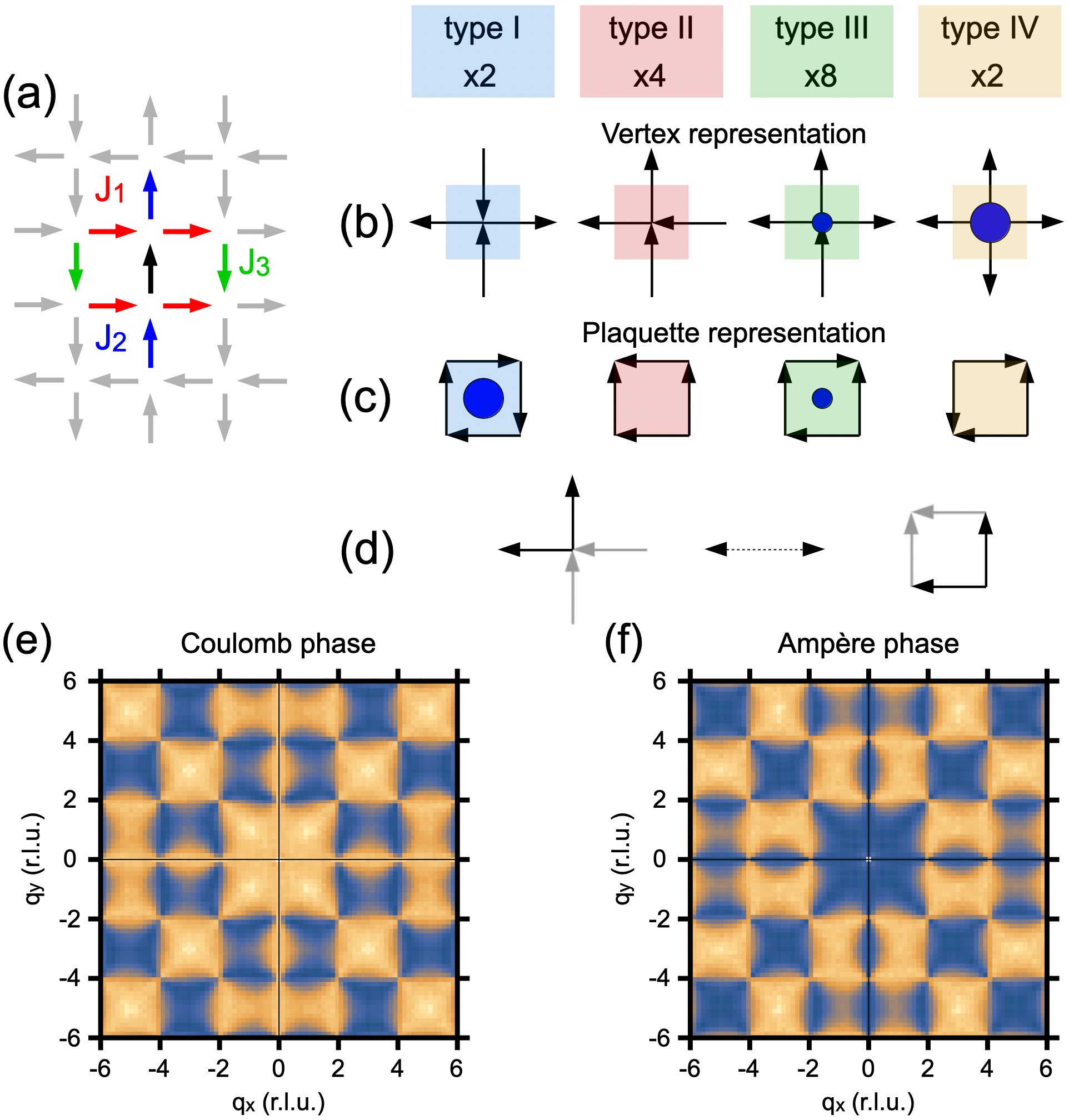}
\caption{(a) Square lattice with Ising spins sitting on the lattice bounds. The three coupling strengths we consider in this work are indicated in red ($J_1$), blue ($J_2$) and green ($J_3$). (b) In the conventional description of the sixteen vertex model, the building block is a cross of four spins meeting at the vertex center. (c) The modified version of the sixteen vertex model consists of four spins surrounding the vertex center. In (a,b) the blue dot highlight the vertices with nonzero divergence or curl. (d) The geometrical transformation allowing to change representation is a translation of a spin pair along the (1 -1) direction, as exemplified for a type-II vertex. The degeneracy of each vertex type is indicated.(e,f) Magnetic structure factors of the Coulomb (divergence-free) (e) and Ampère (curl-free) (f) phases.}
\label{fig1}
\end{figure}

\section{How can we realize an Ampère phase?}
We first consider an Ising spin system on a 2D square lattice, in which the Ising degree of freedom lies on the lattice bonds [see Fig.~\ref{fig1}(a)].
This geometry is interesting as the thermodynamic properties of the Ampère phase can be entirely deduced from those of the seminal square ice, which is a Coulomb phase \cite{Perrin2016}.
To do so, we recall that the square ice is a spin model with ferromagnetic nearest-neighbor interactions only, satisfying the $J_1= J_2$ condition [see Fig.~\ref{fig1}(a)].
Each vertex hosts one of the sixteen possible spin states.
These sixteen configurations can be sorted in four vertex types according to their symmetry [see Fig.~\ref{fig1}(b)], each with energy: 
\begin{equation}
\begin{gathered}
E_\mathrm{I} = - 4J_1 + 2J_2 = - 2J_2~~~~~~
E_\mathrm{II} = - 2J_2\\
E_\mathrm{III} = 0~~~~~~
E_\mathrm{IV} = 4J_1 + 2J_2 = 6J_2,
\end{gathered}
\label{Eq1}
\end{equation}

\noindent with $J_1 = J_2 > 0$.
The Coulomb phase is thus a disordered, yet constrained, tessellation of type-I and type-II vertices.

To describe the curl-free Ampère phase, we now consider a modified model in which the relevant unit is not a \textit{vertex} but rather a \textit{plaquette} [see Fig.~\ref{fig1}(c)]. 
Going from one representation to the other simply requires to translate a pair of spins along the [1 -1] direction, as schematized in Fig.~\ref{fig1}(d).
The same color code is used to represent the equivalence between vertices and plaquettes.
The type-I and type-IV vertices, which are divergence-free (blue) and divergence-full (yellow), respectively, become curl-full and curl-free plaquettes after the geometrical transformation.
Type-III vertices (green), which have a nonzero divergence have a nonzero curl after the [1 -1] translation.
Finally, type-II vertices (red) are divergence-free and transform into curl-free plaquettes.
Key for this work, the energies of the four vertex / plaquette types are identical, provided that the $J_2$ coupling strength in the conventional sixteen vertex model is replaced by $J_3$ in the plaquette representation [see Fig.~\ref{fig1}(a)]. 

However, one difference may come to mind: when describing a square ice system, the divergence-full type-IV vertex has the highest energy, whereas the curl-full plaquette in the modified representation has the lowest energy (at least if one simply replaces $J_2$ with $J_3$ in Eq.~\ref{Eq1}).
To have a complete correspondence between the vertex and plaquettes models, the curl-full type-I plaquette should be the highest energy configuration. 
This could be achieved by exchanging, in the energy hierarchy, this curl-full plaquette with the curl-free (yellow) one [see Figs.~\ref{fig1}(b) and (c)], simply by changing the sign of the $J_1$ coupling strength (see Eq.~\ref{Eq1}).
The highest-energy excitation then corresponds to a full breaking of the divergence-free condition in the square ice model and to a full breaking of the curl-free condition in the plaquette model.
Energy and degeneracy of the two models being exactly the same, their thermodynamic properties map one another.

\begin{figure}
\includegraphics[width=8 cm]{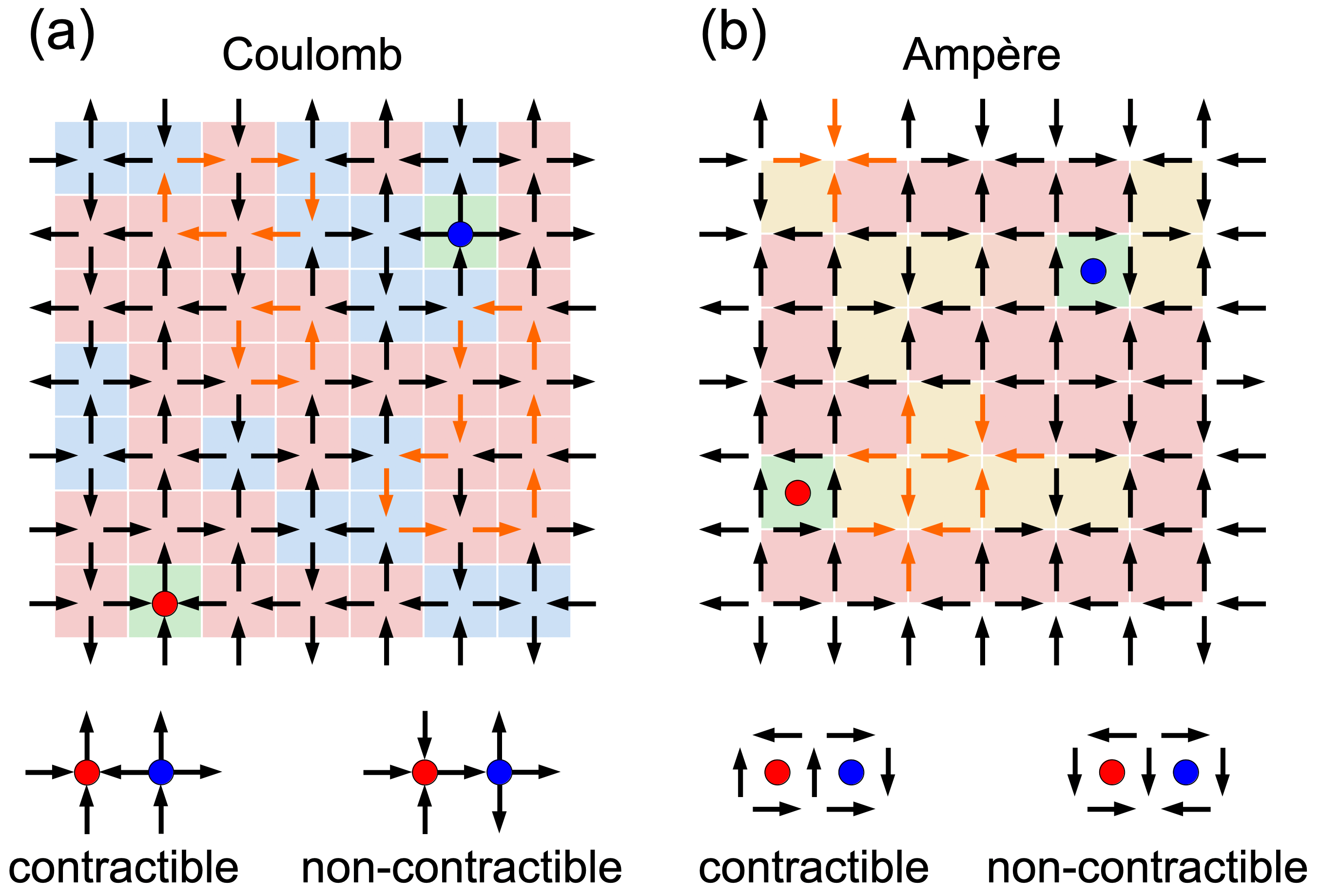}
\caption{(a) Example of a spin configuration belonging to the Coulomb phase and (b) to the Ampère phase. The relevant spin dynamics is illustrated by the orange spins that need to be flipped simultaneously to preserve the divergence-free (a) or the curl-free (b) character of the associated manifold. In the Coulomb phase, the relevant collective motion of spins is a loop dynamics, whereas it is an all in / all out cluster dynamics in the Ampère phase. Topological defects are represented as red and blue dots. They correspond to monopole / antimonopole pairs in the Coulomb phase and to positive / negative currents flowing in a fictional line associated to the local magnetization curl. In both phases, a pair of opposite topological defects can be either contractible or non-contractible, as illustrated by the sketches.}
\label{fig2}
\end{figure}

\begin{figure*}
\includegraphics[width=16 cm]{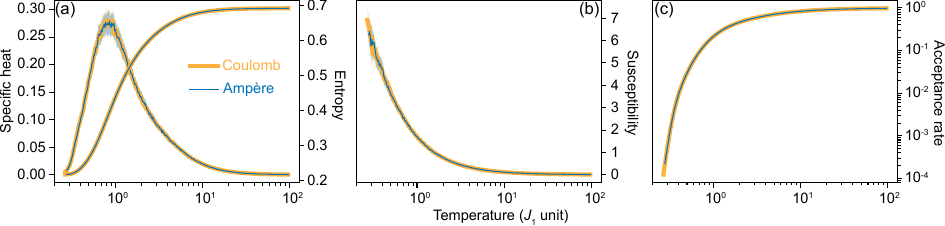}
\caption{\johann{Thermodynamic properties of a 2D Ampère phase, compared to those of a 2D Coulomb phase, computed as function of temperature $T$ with Monte Carlo simulations. (a) Specific heat (left axis) and entropy per site (right axis). (b) Magnetic susceptibility. (c) Acceptance rate during the Monte Carlo simulations. For thermalization, 2$\times$10$^{4}$ modified Monte Carlo steps were used, and the data points are computed with 10$^5$ additional modified Monte Carlo steps (each modified Monte Carlo step consists in multiple local updates, such that stochastic decorrelation is achieved, as checked by on-the-fly calculation of spin-spin autocorrelation times). The procedure whereby the system is cooled down uses logarithmic decrements of 1.02 until the dynamics freezes (corresponding to an acceptance ratio of 5$\times$10$^{-4}$).}}
\label{figThermo}
\end{figure*}

\section{The Ampère phase in a 2D square lattice}
From the above arguments, the spin model which captures the physics of an Ampère phase becomes evident: it is a $\{J_1, J_3\}$ model satisfying the $J_1 =$ $-J_3 = -1$ condition, ensuring $E_\mathrm{II} = E_\mathrm{IV}$.
\johann{We implemented these conditions in Monte Carlo simulations using a spin Hamiltonian of the form}
\begin{equation}
\mathcal{H}=-J_1\sum_{\langle ij\rangle} \sigma_i\sigma_j -J_3\sum_{\langle\langle\langle ij\rangle\rangle\rangle} \sigma_i\sigma_j,
\label{eqHampere2D}
\end{equation}
\johann{\noindent with $\sigma_{i,j}$ Ising variables residing on sites $i,j$, and $\langle\cdot\rangle$ and $\langle\langle\langle\cdot\rangle\rangle\rangle$ denoting pairs of first- and third-nearest neighbors, respectively. For comparison purposes we also consider the Hamiltonian describing a Coulomb (rather than Ampère) phase, \textit{i.e.,}}
\begin{equation}
\mathcal{H}=-J_1\sum_{\langle ij\rangle} \sigma_i\sigma_j -J_2\sum_{\langle\langle ij\rangle\rangle} \sigma_i\sigma_j.
\label{eqHcoulomb2D}
\end{equation}

\noindent Note the difference between the two Hamiltonians: whereas the first one is a \{$J_1, J_3$\} model, the second is a \{$J_1, J_2$\} model.
\johann{The simulations rely on a single spin flip algorithm and were performed for lattices comprising 800 spins ($20\times20\times2$ lattices), in open boundary conditions. 
The specific heat $C$, entropy per site $s$, and magnetic susceptibility $\chi$, were computed as the system was cooled down from a high temperature $T=100\times J_1$.}

\johann{As seen in Fig.~\ref{figThermo}, the three above-mentioned thermodynamic quantities behave identically for the Ampère and the Coulomb Hamiltonians, with a crossover, a residual entropy of about 0.22 at low temperature\cite{Lieb1967a}, and a magnetic susceptibility varying like $1/T$.}

We stress that if the Coulomb phase and its Ampère counterpart share the same thermodynamic properties, the nature of the manifold is strikingly different.
In particular, the magnetic correlations in reciprocal space are distinct, but complementary.
Monte Carlo simulations confirm the analytical predictions developed in Sect.~III, and the magnetic structure factors of the Ampère and Coulomb phases are indeed complementary [see Figs.~\ref{fig1}(e) and \ref{fig1}(f)].

Interestingly, the Ampère phase also suffers from a slowing down of the single spin flip dynamics when approaching the ground state manifold \johann{[Fig.~\ref{figThermo}(c)]}.
This dynamical freezing should be overcome by choosing an appropriate spin update, but the Ampère phase is not a loop model like the Coulomb phase.
Comparing the dynamics in the two phases is instructive to shed light on the nature of the collective spin update needed to further correlate the system.
In a Coulomb phase, the relevant spin dynamics at low temperature is a loop dynamics [see Fig.~\ref{fig2}(a)].
This is so because the Coulomb phase is a divergence-free manifold, and a collective spin update must preserve this divergence-free character.
It is thus a curl-full spin update, i.e., a loop move.
In contrast, an Ampère phase is a curl-free manifold, and the relevant collective dynamics which preserves this curl-free character is a divergence-full spin update.
The only spin flip dynamics which further correlates the Ampère phase when reducing the temperature is a complete reversal of all in / all out vertices, as illustrated in Fig.~\ref{fig2}(b).

Within a single spin flip approach, the lattice dynamics is entirely determined by the motion of topological defects.
In the Ampère phase, these defects are type-III plaquettes carrying a partial curl of magnetization.
Flipping one spin in an Ampère phase leads to the formation of a pair of topological defects having opposite chirality.
This chirality is the equivalent of the monopole charge in the Coulomb phase.
We note that the correspondence between monopoles and current lines is valid for both contractible or non-contractible pairs (see Fig.~\ref{fig2}).
Depending on the orientation of the shared spin, two oppositely charged defects can either recombine upon reversal (contractible pair) or be trapped to avoid the formation of a high-energy, doubly charged pair (non-contractible pair).
Therefore, the relaxation of topological defects in the Ampère phase is expected to occur at different time scales, like in the Coulomb phase \cite{Castelnovo2010}, when dipolar interactions are present in the system.

\begin{figure}[htp]
\centering
\includegraphics[width=8 cm]{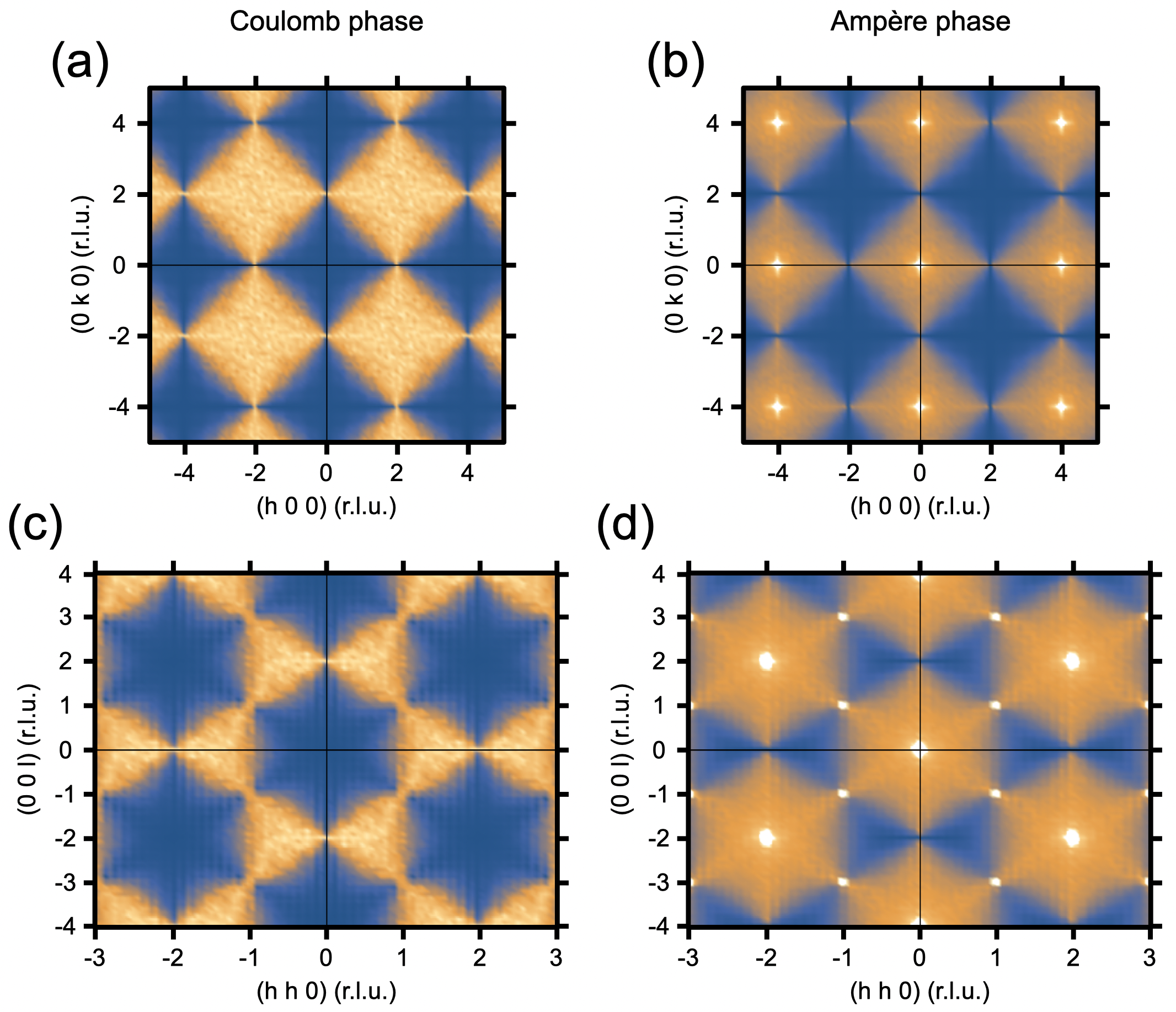}
\caption{\Nico{Magnetic structure factors within two different crystallographic planes computed for a divergence-free (a,c, Coulomb phase) and a curl-free (b,d, Ampère phase) constraint on a pyrochlore lattice. \new{To ensure independence of measurements, each one is done after one modified Monte Carlo step, which consists in $N$ cluster updates ($N$ being the number of sites). Because the loop search is more time consuming, $5 \times 10^5$ of such steps have been done, while $3 \times 10^7$ have been performed in the curl-free case.}}}
\label{fig3}
\end{figure}

\section{The Ampère phase in 3D}

The curl of a coplanar vector field has one component only, and in that case the curl-free constraint is equivalent to the divergence-free constraint provided that the field is rotated by $\pi/2$.
Considering this simple transformation, the thermodynamics as well as the coarse-grained field properties of the two models are the same.
However, identifying a field transformation is not always so straightforward. 
In the general case, there is indeed no simple mapping between the $\vec{\nabla}\cdot$ and $\vec{\nabla}\times$ operators, i.e., establishing the connexion between the thermodynamics of the Ampère and Coulomb phases is not trivial. 
In the following, we thus focus on another seminal frustrated spin lattice, the 3D pyrochlore spin ice \cite{Note1}, and compare their curl-free and divergence-free constrained ground state manifolds.

\noindent To do so, we identified a stochastic dynamics that efficiently samples the manifold.
In the divergence-free scenario, it is the well-known loop dynamics.
In the curl-free case, we found that efficient spin updates consist in the reversal of full tetrahedra, whose spins all point either in or out (reminiscent of the update we identified in the 2D square lattice).
While such a spin update preserves the curl-free constraint, it probably suffers from ergodicity breaking as it does not change the number of all-in and all-out tetrahedra.
To avoid this bias, we first prepared a set of micro-states from the ground state manifold.
With these states at hand, we then implemented the cluster dynamics to probe each 'sector' defined by an initial micro-state.

\begin{table}
\caption{\label{table1} \new{Pairwise spin correlations as a function of distance for the Coulomb and Ampère phases in the pyrochlore lattice.}}
\begin{tabular}{|p{1cm}|p{2.2cm}|p{2.2cm}|}
\hline
\hline
 $n$ & $C_{n}$ (Coulomb) & $C_{n}$ (Ampère) \\
\hline
1 & 0.333385 & 0.355972  \\
2 & 0.0422152 & 0.047282  \\
3 & 0.00947359 & 0.0139193  \\
4 & 0.00328901 & 0.00739311  \\
5 & 0.00148869 & 0.00533742  \\
6 & 0.00101538 & 0.00448053  \\
7 & 0.000545788 & 0.00404486  \\
8 & 0.000216199 & 0.00381041  \\
9 & 0.00029943 & 0.0036741  \\
10 & -- & 0.00362052  \\
\hline
\hline
\end{tabular}
\end{table}

\new{Monte Carlo simulations were performed under periodic boundary conditions for a $9\times9\times9\times16$ lattice for the divergence-free constraint, and a $10\times10\times10\times16$ for the curl-free case.
Note that referring to our simulations as Monte Carlo ones may be confusing: we are here dealing with a zero temperature dynamics. 
Each proposed update is accepted as we are probing the constant energy manifold of each model ground state.  
Correlations are taken along lines defined by nearest neighbors of the pyrochlore lattice. 
To deal with sites of the same given type (defined by their local anisotropy axis), correlations reported in Table~\ref{table1} are those involving an arbitrary central site and neighbors of the same type along a line, i.e., separated by a distance of two nearest neighbors distance.}

The magnetic structure factors, computed for the Coulomb and Ampère phases, are here also complementary (Fig.~\ref{fig3}), consistent with our analytical description of the magnetic correlations (see also Ref. \onlinecite{Gingras2023}).
Importantly, in both cases we spot the presence of pinch points, i.e., the reciprocal space translation of an algebraic decay.
To tell whether the phases are Coulombic or Amperian, we need to determine the algebraic exponent of the spin-spin decay over distance.
\new{We indeed find correlation laws in the form $a/r^b$ with $b = 3.041 \pm 0.042$ in the Coulombic case ($a = 0.333 \pm 0.001$), and $c+a/r^b$ with $b = 3.018 \pm 0.030$ in the Amperian case ($a = 0.353 \pm 0.001$ and $c = 0.0028 \pm 0.0004$).}

\Nico{Two remarks should be made.
First, both phase exponents are consistent with their space dimensionality, which legitimates our interpretation.
This aspect is far from trivial as it is related to the {\it hypothesis} that once the central limit theorem is applied on a coarse-grained part of the system, the density of magnetization assumes a normal distribution, as for a true paramagnet.
Second, we find a constant contribution to the spin-spin correlations in the Ampère phase.
It is this component that gives rise to emerging Bragg peaks in the magnetic structure factors [Figs.~\ref{fig3}(b) and \ref{fig3}(d)].
Unexpectedly, we observe that this peculiar Ampère phase fragments, as Coulomb phases do \cite{Brooks2014, Jaubert2015} in other situations \cite{Canals2016, Petit2016, Paddison2016, Elsa2020}.
}

\textit{Prospects.--}
\Nico{The concept of Ampère phase broadens the analogy developed between highly frustrated magnets and electromagnetism, and opens new avenues for theoretical and experimental investigations. 
\new{In particular, the Ampère phase with its curl-free constraint is not described so far in the recent attempts to characterize and classify spin liquids \cite{Benton2021, Pujol2023, Benton2024, Moessner2024, Gingras2024}.}
Like Coulomb phases are found in many geometries (square, kagomé, pyrochlore), Ampère phases are likely ubiquitous as long as the underlying physics provides with cooperative paramagnetism and a local, curl-free conservation law.
As observed for instance in our 3D example, these phases are also subject to fragmentation.
While unanticipated, this is not totally surprising.
Indeed, fragmentation corresponds to two (almost) independent behaviors of the divergence-free and curl-free components emerging from a Helmholtz decomposition \cite{Brooks2014}.
This effective independence, known to occur in Coulomb phases, occurs as well in Ampère phases, as our 3D example shows.
Future theoretical investigations will hopefully deepen the understanding of such an exotic ground state, and extend the concept of Ampère phase beyond the case of Ising spins --- towards $d$-algebraic spin liquids with continuous spin degrees of freedom \cite{Canals1999,Canals2001,Canals2002} where true dynamical effects (classical \cite{Robert2008} or quantum \cite{Gingras2023}) become relevant.
Artificial spin systems \cite{Nisoli2013, Rougemaille2019, Heyderman2020} might also be an interesting route to follow to visualize 2D Ampère phases in real space and time. 
}


%

\end{document}